# Coulomb Crystallization of Highly Charged Ni$^{12+}$ Ions in a Linear Paul Trap

Shaolong Chen[1, #], Zhiqiang Zhou[1,2,#], Guosheng Zhang[1,2], Jun Xiao[3], Yao Huang[1], Kelin Gao[1,4,*], Hua Guan[1,4,5,*]

**AFFILIATIONS**

[1]State Key Laboratory of Magnetic Resonance and Atomic and Molecular Physics, Innovation Academy for Precision Measurement Science and Technology, Chinese Academy of Sciences, Wuhan 430071, China.

[2]University of Chinese Academy of Sciences, Beijing 100049, China

[3]Shanghai EBIT Laboratory, Key Laboratory of Nuclear Physics and Ion-Beam Application (MOE), Institute of Modern Physics, Fudan University, Shanghai 200433, China

[4]Hefei National Laboratory, Hefei 230088, China

[5]Wuhan Institute of Quantum Technology, Wuhan 430206, China

[#)] **These authors contributed equally to this work.**

[*)] **Authors to whom correspondence should be addressed:** klgao@apm.ac.cn and guanhua@apm.ac.cn.

## ABSTRACT

Optical clocks have garnered widespread attention due to their unparalleled precision in time-frequency standards, geodetic measurements, and fundamental physics research. Among emerging developments, highly charged ion (HCI)-based optical clocks have attracted significant scientific interest owing to their exceptional resilience against electromagnetic perturbations and enhanced sensitivity to variations in the fine-structure constant ($\alpha$). While the recent successful demonstration of an Ar$^{13+}$ optical clock has validated the feasibility of HCI-based systems, Ni$^{12+}$—featuring an ultranarrow clock transition linewidth—stands out as a superior candidate for achieving HCI optical clocks with $10^{-19}$ level uncertainty and stability. In this work, we report the Coulomb crystallization of nickel highly charged ions (Ni-HCIs). Through a precision deceleration and sympathetic cooling protocol in a room-temperature Paul trap, high-energy Ni-HCI bunches were sympathetically cooled from megakelvin to the 100-millikelvin range using laser-cooled Be$^+$ ions. This work represents a pivotal step toward the realization of optical clock based on Ni$^{12+}$ ion.

## I. INTRODUCTION

HCIs possess clock transitions with intrinsic accuracies below $10^{-19}$ [1]. Notably, these transitions exhibit high sensitivity to variations in the fine-structure constant α, making HCIs a valuable platform for exploring physics beyond the Standard Model [2–5]. Due to their unique properties, HCIs have attracted significant interest from researchers in recent years. Over the past few decades, optical clock precision has continuously improved [6–13]. However, further reducing uncertainties has become increasingly challenging due to the fundamental physical properties of atomic or singly charged ion transitions. A promising approach to overcoming this limitation is to utilize transitions in HCIs or atomic nuclei as frequency references. This is because such transitions experience stronger interactions from nuclei, rendering them highly resistant to external perturbations while simultaneously exhibiting exceptional sensitivity to variations in the fine-structure constant. As a result, transitions of HCIs and nuclear are among the most promising candidates for pushing the limits of optical clock precision and enabling new tests of fundamental physics.

Thanks to the relentless efforts of research groups worldwide, significant experimental progress has recently been achieved. For example, the precision of nuclear transitions in thorium-229 ($^{229}$Th) has been continuously advanced [14–20], and the transition was recently measured with high precision [21–23]. Concurrently, more implementation schemes for nuclear optical clocks have been proposed [24,25], sparking widespread interest in this field. On the other hand, research on HCI optical clocks has been continuously advancing, particularly regarding studies related to $Ar^{13+}$, From initial measurements of clock transition to the development of cryogenic ion trap systems [26–30], ion trapping and cooling [31,32], and logic spectroscopy measurements [33], significant progress has been achieved. Currently, the $Ar^{13+}$ optical clock has realized system locking and evaluations of uncertainty and stability. An HCI optical clock with an uncertainty of $2.2\times10^{-17}$ has been demonstrated for the first time [34] by Piet O. Schmidt's group. This breakthrough confirms the applicability of HCIs in optical clocks.

However, while $Ar^{13+}$ was the first HCI system to be implemented as an optical clock, it also has notable drawbacks, particularly the large natural linewidth of its clock transition, which limits interrogation times and its potential for achieving instability at the $10^{-18}$ or even $10^{-19}$ level. Consequently, identifying more suitable HCI systems for optical clock applications has become a major research focus. On the theoretical front, efforts to screen candidate HCI ions for optical clocks have been ongoing. Numerous candidate HCI optical clock systems have been proposed, such as $Cf^{15+,17+}$ [35], $Ho^{14+}$ [36], $Ir^{17+}$ [37,38], $Os^{16+}$ [38], $Ni^{12+}$ [1], $Pr^{10+}$ [39]and others [4,37,40–46]. In parallel, spectroscopic measurements and related investigations for some of these ions are being actively pursued [47–50]. Our team has been actively investigating an optical clock based on Ni-HCIs. Several highly charged states of nickel exhibit suitable clock transitions, with $Ni^{12+}$ being particularly promising [1]. This ion features two potential clock transitions: one electric quadrupole (E2) transition with an exceptionally narrow natural linewidth of approximately 8 mHz, making it an ideal candidate for achieving ultrahigh precision, and one magnetic dipole (M1) transition with a natural linewidth of approximately 24 Hz, which can serve as a secondary clock transition or as an auxiliary transition to enhance the detection efficiency of the E2 transition. Recent theoretical studies have extensively analyzed the transition wavelengths and clock properties of $Ni^{12+}$. These analyses indicate that, under current experimental conditions, systematic shifts such as electric quadrupole shifts and second-order Zeeman effects can be controlled to below the $10^{-19}$ level [1,51].

In the field of HCI trapping and cooling, a breakthrough was achieved in 2015 when J. R. Crespo López-Urrutia's team demonstrated the sympathetic cooling of $Ar^{13+}$ ions in a cryogenic ion trap [31], and more recently, they have also reported the cooling of Xe-HCI [32], overcoming a key technical bottleneck. Although there had been earlier demonstrations of cooling for a few other ion species, such as $Th^{3+}$ [52] and $Ca^{2+}$ [53], and more recently for sympathetically cooled $Th^{3+}$ [54], reports on the cooling of HCIs with charge states above +5 remain scarce. This is primarily because HCIs are typically produced via high-energy bombardment methods like accelerator, Electron Cyclotron Resonance source (ECR), or Electron Beam Ion Trap (EBIT), which generate ions with high initial kinetic energy, and require beamline to transfer ions into Paul traps for subsequent trapping and cooling to millikelvin temperatures. Consequently, the trapping and cooling of HCIs has remained one of the key technical challenges in HCI optical clock research, yet the search continues for HCI systems more suitable for breaking the current limits of optical clock precision, such as $Ni^{12+}$, $Cf^{17+}$, among others.

In our studies, we initiated investigations of $Ni^{12+}$ by constructing a superconducting electron beam ion trap (SW-EBIT) [55], producing $Ni^{12+}$ ions, and measuring the M1 clock transition within the EBIT [48,56]. However, several major challenges remain in the realization of a $Ni^{12+}$ optical clock, among which ion trapping and cooling represent critical aspects. Here, we implement sympathetic cooling of Ni-HCIs using $Be^+$ ions in a room-temperature ion trap. We first prepare a large ensemble of laser-cooled $Be^+$ ions in the trap and then inject Ni-HCIs produced in the EBIT into the same trap [57]. Through Coulomb interactions between $Be^+$ and the HCIs, we achieve effective cooling of the $Ni^{12+}$, reducing their temperature by seven orders of magnitude from the megakelvin range down to the 100 mK level.

## II. Experimental setup

The Ni-HCIs were produced using the SW-EBIT [55], from which ion bunches were extracted in pulsed mode with an energy of approximately 700 qV. The time-of-flight (TOF) method was employed to distinguish the charge state of the HCIs, and pure $^{58}Ni^{12+}$ ions were selected by applying a pulsed voltage to a parallel electrode plate. The filtered ions were then decelerated and injected into the ion trap through the combined action of a rapidly switched electric field and electrostatic lenses. Concurrently, $^9Be^+$ crystal was first prepared within the ion trap. As the $^{58}Ni^{12+}$ ions were injected, coulomb interactions between the $^9Be^+$ and $^{58}Ni^{12+}$ ions progressively cooled the $^{58}Ni^{12+}$ ions.

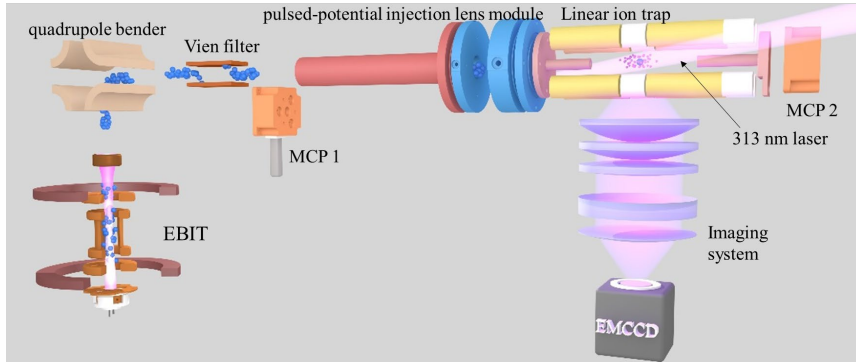

**Fig.1.** Schematic diagram of the complete experimental setup: Ni-HCI from production and deceleration to trapping and cooling.

HCIs, which require stripping multiple electrons and demand high ionization energies, led us to select an EBIT as the ion source here. By optimizing the EBIT parameters, we maximized the proportion of $Ni^{12+}$ ions in the ion beam. The ion bunches were extracted in pulsed mode with a pulse interval of 100 ms and a pulse width of approximately 10 μs. The ion beam direction was redirected from vertically upward to horizontally rightward using a quadrupole bender. Following this, a Wien filter was employed for preliminary charge-state selection and analysis. However, due to limitations in the Wien filter's magnetic field strength, flight region geometry, and aperture size, its resolution proved insufficient to distinguish ions of the same charge state but different isotopes (e.g., isotopic variants of Ni$^{12+}$). To address this, we removed the Wien filter's magnetic field and repurposed it as a parallel electrode plate system. By applying a pulsed deflection voltage to the plates synchronized with the time-of-flight (TOF) measurements, we effective filtered both charge-states and isotopes. For further details on this process, refer to Ref. [58].

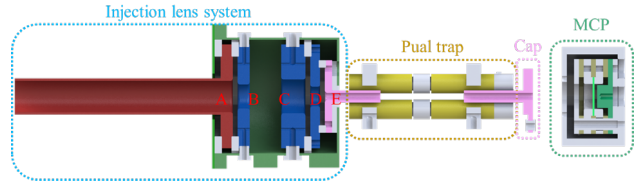

**FIG. 2.** Overview of the beamline for the deceleration, injection, and trapping of HCIs. The beamline consists of pulsed-potential injection lens module, a Paul trap, and a Microchannel plate detector (MCP).

Well, the kinetic energy of filtered HCIs is too high to be trapped in a typical Paul trap. Here we have developed a pulsed-potential injection lens module. This module comprises five cylindrical stainless-steel electrodes housed within a shielded stainless-steel cylinder. The energy reduction process initiates at drift tube A (13 mm inner diameter), inside is an equipotential region. Upon ion bunch entry, the entire tube's potential is rapidly pulsed to a negative value within several nanoseconds. While this transition preserves the ion's kinetic energy, it effectively reduces their potential energy. Subsequent exit from this region through an ascending potential gradient, their kinetic energy is partially converted to potential energy, achieving preliminary deceleration to approximately 100 qV.

Following this deceleration stage, the beam passes through a three-element electrostatic lens assembly comprising electrodes B, C, and D with respective inner diameters of 10 mm, 14mm, and 8mm. Those lenses counteract beam divergence induced by the deceleration. Simulations confirm the effectiveness of lenses in beam collimation and transmission enhancement, as show in Fig. 3, the blue trajectories.

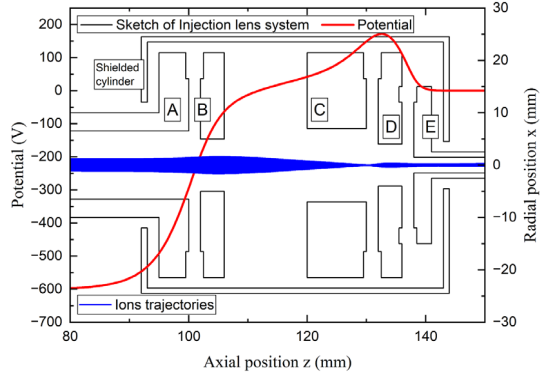

**FIG. 3.** Schematic of injection lens system. The simulated trajectory of HCIs across the injection lens system is shown by the blue lines. The gray lines show the sketch of the injection lens system, which consists of a drift tube A, electrostatic focusing lens group B, C, and D are enclosed in a shielding cylinder. The red line presents the distribution of potential. The HCIs decelerated from ~700 qV to ~100 qV, and then were collimated by the electrostatic lens module.

At the end of the beamline is the Paul trap, and Electrode E and the Cap (both with 3mm inner diameters) are mirror-symmetrically positioned at both axial ends of the Paul trap. These electrodes protrude approximately 10 mm into the quadrupole rod structure. Upon HCIs entry into the trap, the voltages on these electrodes are rapidly elevated to +150 V within nanosecond timescales to establish axial confinement potentials, working synergistically with the radio-frequency quadrupole field to achieve three-dimensional confinement of the HCIs. In the beamline, MCP detectors were installed before the deceleration drift tube and after the ion trap. The MCP before the deceleration drift tube could be manually retracted from the beamline, while the MCP after the ion trap was fixed in place. The MCP detectors were used to measure the ion state parameters, including the arrival time of the ions at the detector and the ion beam intensity.

## III. Ion trapping and cooling

In our experiment, Ni-HCIs were extracted in pulsed mode by raising the voltage of the central electrode in the EBIT trapping region. For the three-segment EBIT trap, the potentials of the end electrodes were set such that the voltage on the electron gun side was slightly higher than that on the extraction side, while both remained lower than the raised voltage of the central electrode but higher than its initial voltage before being raised. Additionally, the time point at which the trap voltage was raised was defined as the zero-signal trigger time, serving as the reference for timing the operations of all subsequent electrodes involved in voltage modulation. Throughout the entire ion beamline, variable voltages were applied to several components, including the Wien Filter for ion selection, the tube for deceleration, and the mirror electrodes at both ends of the ion trap for ion injection.

Fig. 4 presents the voltage values and timing sequences of all electrodes requiring pulsed modulation, where $t_0$ represents the initial time of ion extraction. At time $t_0$, the voltage on the central electrode of the EBIT trap region is raised from $U_0$ to $U_{ex}$ (approximately 720 eV), initiating ion extraction, The interval between two extractions is T, approximately 100 ms. Monitoring results indicate that at this extraction frequency, the production yield of $Ni^{12+}$ ions is maximized. The extraction pulse duration $T_1$ is about 10 us. After $t_1$, the ions free flight and reach the commercial Wien Filter (DREBIT), and the target ions with the desired mass-to-charge ratio are selectively filtered out based on the potential difference applied across the parallel plate electrodes of the Wien Filter (For a detailed description of the process, please refer to Ref. [58]). At $t_2$, after all Ni-HCIs beam transmitted through the beamline into the drift tube A, we quickly drop the potential of the A to the negative $-U_{low}$ within 40 ns (employ a circuit based on a commercial high-speed high-voltage switch (HTS 41-03, Behlke)) to rapidly reduce the ions' potential energy. At $t_3$, the ions arrive at the entrance of the ion trap, where the voltage on electrode E decelerate from initial high voltage $U_{end}$ to $U_{bias}$ to facilitate ion entry. Subsequently, the potential is raised back to $U_{end}$, achieving the confinement of HCIs.

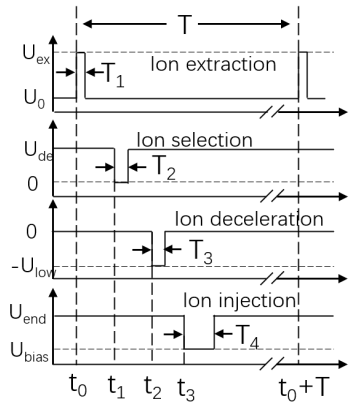

**FIG. 4.** Time sequences of all electrodes requiring pulsed modulation.

To enhance the probability of HCIs being successfully injected into the Paul trap, we did not directly decelerate the kinetic energy to below 10 qV, instead, we employed a two-step deceleration approach. The initially decelerated ions were slowed to approximately 100 qV using the injection lens system, while the secondary deceleration of ions during the injection phase is accomplished using an ion trap. We raised the potential of the ion trap to approximately 100 V. The initially decelerated ions are axially injected into the ion trap through E, which is initially maintained at a high potential of approximately 150 V to prevent contaminant ions from entering the trap. However, at this stage, the kinetic energy of the decelerated ions is lower than the E potential barrier, preventing them from entering the ion trap. And the pulse width of the HCI beam is less than 0.5 μs, the experiment employs precise timing control at t₃ to open an injection window T₄ of about 1 μs in E. This ensures accurate injection of HCIs into the ion trap while effectively suppressing the entry of contaminant ions. As the ions entered the Paul trap, they ascended the potential gradient once more, undergoing a second stage of deceleration and ultimately meeting the trapping conditions.

Like most other HCIs, $^{58}$Ni$^{12+}$ lacks a suitable transition for laser cooling. Therefore, both cooling and subsequent spectroscopic measurements require the involvement of auxiliary ions. In this work, $^9$Be$^+$ ions were used to sympathetically cool Ni$^{12+}$ and serve as the logic ions for quantum logic spectroscopy. The primary reason for choosing Be$^+$ is that its charge-to-mass ratio is relatively close to that of $^{58}$Ni$^{12+}$.

Experimentally, ion trapping is performed employ a linear Paul trap [59], and anti-phase radio frequency signals ($V_{RF}(t)$: $f_{RF} = 9.718$ MHz with several hundred volt) are loaded on two pairs of electrodes in the trap, the end cap voltage of Paul trap ($V_{End}$) is about 1 V to 5 V, as shown in Fig. 5(a). Ion crystal with hundreds Be$^+$ was first prepared in the trap. We use a 523 nm Q-switched laser (Changchun New Industries Optoelectronics Tech. Co., Ltd. (CNI), EL-532-1.5W) to sputter a Be metal target to generate $^9$Be$^+$. As shown in Fig. 5(b). The ground state ($^2S_{1/2}, F = 2, m_F = 2$), ($^2S_{1/2}, F = 1, m_F = 1$) and the excited state ($^2P_{3/2}$) form a cooling cycle. The typically hyperfine structure splitting of the ground state is 1.25 GHz, so we built a double-pass optical path based on AOM to achieve a frequency shift of 1.25 GHz. The cooling and repump laser are then overlapped using a beam splitter and directed through the ion trap. The fluorescence signal of the ion crystal is imaged onto an EMCCD (Andor iXon Ultra DU-888) using a UV imaging lens with a numerical aperture of 0.32 and a working distance of 70 mm, positioned outside the vacuum system.

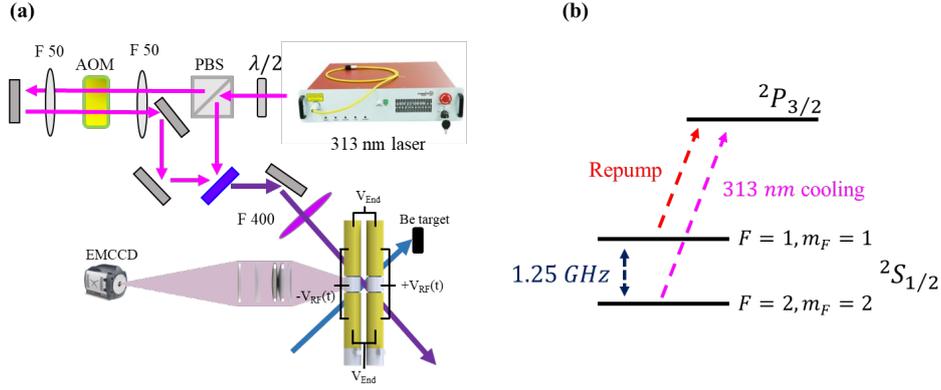

**FIG. 5.** The ion trapping system and energy levels of $^9Be^+$.

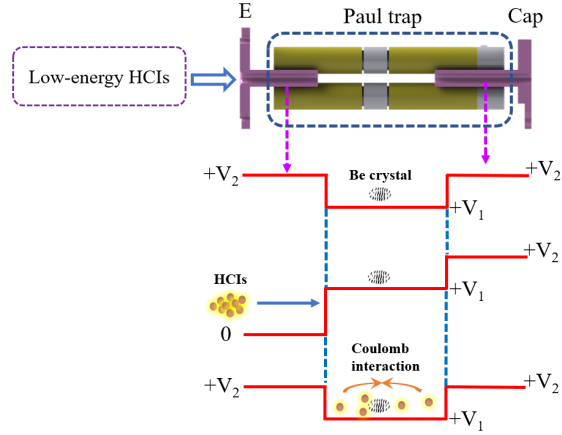

**FIG. 6.** Voltage sequence of HCIs injection. Beryllium ion crystal is prepared in Paul trap first, and then the potential of E is dropped to inject HCIs.

The observed Be$^+$ crystals showed no significant changes when the voltages of E and Cap were raised simultaneously. Additionally, when Cap was first set to 150 V and E was subsequently raised to 150 V, the ion crystal shifted only about 10 μm to the right, without affecting its internal structure. E is initially held at a high potential to prevent unwanted ions from entering the ion trap. For HCIs implantation, the potential of E must be adjusted sequentially to create an injection window. The timing of the entire experiment is controlled by a commercial multi-channel digital delay pulse generator (DDG-210, accuracy 1.25 ns, Becker & Hickl). To determine the optimal injection timing, we first ground the Paul trap and Cap and observe the HCIs TOF signal using the MCP detector. The start times of E and DFA are synchronized, and the low-level duration of E is extended until the ion signal on the MCP remains unchanged. Next, the E start time is delayed until a change in the MCP signal is observed, indicating that the falling-edge potential of E is affecting the HCI bunch. At this point, the timing is finely adjusted back until the MCP signal stabilizes. Then, the duration of E's low-level phase is gradually reduced until the MCP signal changes, indicating that the rising-edge potential of E is influencing the HCI bunch. The low-level duration time is then finely extended until the MCP signal remains stable, thereby determining the optimal timing window and ensuring the successful injection of HCIs into the trap. At this final injection stage, the ion trap potential is raised and maintained at 100 V, while the Cap potential is set to 150 V, several ion pulses are then injected, and after waiting for a few to several tens of seconds, the two-component Coulomb crystallization of Ni-HCIs and $Be^+$ is observed in the Paul trap.

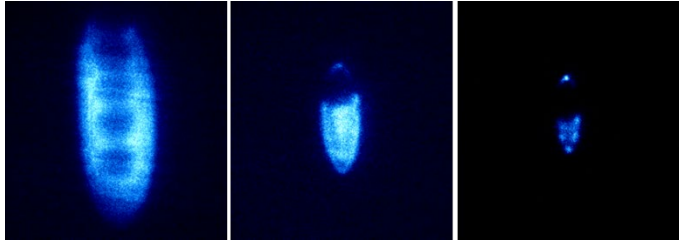

**FIG. 7.** Coulomb crystal of $Be^+$ and $^{58}HCI^{12+}$ (the dark circular shapes).

To better understand the process of sympathetic cooling, we conducted molecular dynamics simulations. In the simulation, we developed a Python-based simulation program for ion trapping and cooling, following the method outlined in Ref. [60]. The ion trap parameters in the simulation were set to match those used in the experiment. By introducing an appropriate random heating force, we generated a series of ion crystal images under different conditions. Comparing the equilibrium-state ion distributions from the simulations with the experimentally observed images enabled us to identify the simulated configurations that best reproduced the experimental results. From these matched simulations, we extracted detailed information on ion trajectories and velocities, allowing us to estimate the temperature and charge state of both the $Be^+$ and HCI ions present in the experiment.

For resolving ionic charge states, we using clear crystallized ion images of the test sympathetic cooling ions before and after the same experimental cycle as calibration references. By comparing the simulated ion crystal dimensions with experimental imaging results, we verified the charge states of the sympathetic cooling ions. The experimental setup utilized an EMCCD with single-pixel resolution of 13 μm and an imaging system magnification of approximately 15.4. For crystallized ion measurement, we defined the experimental crystal size as the center-to-center distance between the two outermost ions along the trap axis. Correspondingly, the simulation calculated each ion's average position after reaching steady-state, with the simulated crystal size determined by the position difference between axial outermost ions in the molecular dynamics model. Through systematic comparison between experimental and simulated crystal dimensions, we validated the simulation system's accuracy. Furthermore, the observed size discrepancies provided quantitative information about the sympathetic cooling ions' charge states.

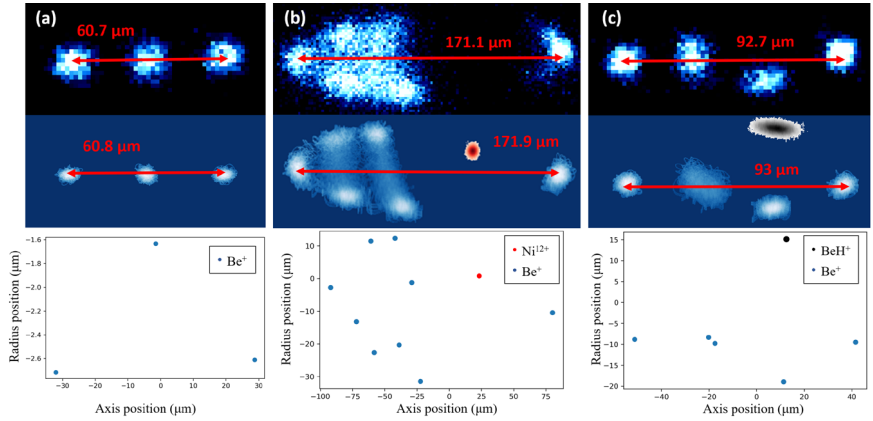

**FIG. 8.** Comparison between experimental and simulated images (top: experimental image, middle: simulated fluorescence image, bottom: equilibrium positions of each ion from simulation). The radio-frequency and static trapping parameters were: trapping RF frequency of 9.718 MHz, peak-to-peak RF voltage of 160 V, and endcap voltage of 1 V. (a) Image of three $^9Be^+$ ion, showing an axial dimension of 60.7 μm, while the molecular dynamics simulation yielded 60.8 μm; (b) Sympathetically cooled ion crystal consisting nine $^9Be^+$ and one $Ni^{12+}$, with an observed axial length of 171.1 μm; the simulated image (with the $Ni^{12+}$ marked in red) shows an axial length of 171.9 μm; (c)Crystal consisting of five $^9Be^+$ and one $BeH^+$, with an axial dimension of 92.7 μm in the experiment and 93 μm in the simulation (with the $BeH^+$ dark ion shown in black).

Fig. 8 presents a comparative analysis of experimental and simulated images. The top images in panels (a), (b), and (c) show experimental results for different ion crystals under identical trapping and cooling parameters. The middle images in each panel display simulations conducted under identical conditions (confinement frequency: 9.718 MHz, peak-to-peak voltage: 160 V, endcap voltage: 1 V), while the bottom images represent the equilibrium-state average ion positions from molecular dynamics simulations. As illustrated in Fig. 8(a), The EMCCD imaging resolved three ions with 72-pixel separation between axial endpoints, corresponding to 60.7 μm (uncertainty: ±1.7 μm, accounting for 2-pixel resolution limit at 13 μm/pixel and 15.4 magnification). Molecular dynamics simulation yielded an axial dimension of 60.8 μm, showing excellent agreement. Fig. 8(c) Crystal composed of five $^9Be^+$ ions and one $BeH^+$ dark ion. Experimental axial measurement (92.7 μm) matched simulated results (93 μm), validating the simulation's predictive capability for co-cooling behavior. The co-cooling regime in Fig. 8(b) involved sympathetically cooled ion crystal consisting of nine $^9Be^+$ and one $Ni^{12+}$, with an observed axial length of 171.1 μm, the simulated image shows an axial length of 171.9 μm. Crucially, substituting $Ni^{12+}$ with $Ni^{11+}$ or $Ni^{13+}$ yielded 167.8 μm and 175.8 μm dimensions respectively, conclusively identifying the co-cooling species as $Ni^{12+}$ through this charge-state fingerprinting method. Furthermore, during the injection process, the charge state of the incoming ions can also be well controlled, which is consistent with the expected charge state, For the temperature, we obtained simulated images that closely matched the experimental results, from the simulation, the temperature evolution curve was extracted, yielding a final equilibrium temperature of 100 mK level. Meanwhile, by selecting different charge states and optimizing the pulse timing of various electrodes, we experimentally achieved sympathetic cooling of HCIs ranging from $Ni^{7+}$ to $Ni^{13+}$, as well as Ar-HCIs, these experiments provided valuable insights into optimizing the overall experimental scheme.

## IV. CONCLUSION

In this paper, we have demonstrated the sympathetic cooling of Ni-HCIs in a linear ion trap. The process began with the extraction of ions from the EBIT, followed by charge-state and isotope selection to obtain $^{58}Ni^{12+}$. The ions were then initially decelerated using a drift tube, further slowed down and injected into the trap through a deceleration lens system, ultimately achieving the confinement of HCI ions. Meanwhile, we selected $^9Be^+$ ions, which have a charge-to-mass ratio close to that of $^{58}Ni^{12+}$, as the laser-cooled ions for sympathetic cooling. As a result, we successfully formed two-component Coulomb crystals of $Ni^{12+}$ and $Be^+$ ions, with ion temperature reaching 100 mK level. This marks a crucial step toward the development of an HCI-based optical clock using $Ni^{12+}$, distinguishing it from traditional single-charged ion or neutral atom optical clocks and addressing one of the key challenges in $Ni^{12+}$ clock development, this work lays the foundation for future research on $Ni^{12+}$ based optical clocks. However, the problem that still exists in this work is that sympathetic cooling was performed in a room-temperature ion trap with a vacuum pressure of only $3\times10^{-8}$ Pa. Due to their high charge state, HCI ions are highly susceptible to electron capture from residual gas collisions [30], limiting their lifetime to only a few tens of seconds. This short trapping duration constrains further experimental operations. Therefore, to achieve a higher vacuum, we are actively developing a cryogenic ion trap system as part of our ongoing efforts.

## ACKNOWLEDGMENTS

We sincerely thank Huanyao Sun for his assistance with the RF circuitry and Shiyong Liang for his contributions to the early-stage work. This work was supported the Innovation program for Quantum Science and Technology (Grant No. 2021ZD0300901), the CAS Project for Young Scientists in Basic Research (Grant Nos. YSBR-085 and YSBR-055), the National Natural Science Foundation of China (Grant No. 92265206, 12393823, 12121004), and Natural Science Foundation of Hubei Province (Grant No. 2022CFA013).